\documentclass[12pt,technote,onecolumn,draftcls]{IEEEtran}

\usepackage[top=1.7cm, left=1.8cm, right=1.8cm, bottom=1.8cm]{geometry}
\usepackage{graphicx}
\usepackage[english]{babel}
\usepackage[utf8x]{inputenc}
\usepackage{amsmath}
\usepackage{amssymb}
\usepackage{pifont}
\usepackage{graphicx}
\usepackage{hyperref}
\usepackage{cite}
\usepackage{xcolor}

\title{Speaker localization using direct path dominance test based on sound field directivity}

\author{Boaz Rafaely and Koby Alhaiany\thanks{This work was supported by the Israel Science Foundation (ISF) under Grant 146/13.} \thanks{The authors are with the Department of Electrical and Computer Engineering, Ben-Gurion University of the Negev, Israel} \thanks{Corresponding email: br@bgu.ac.il}}

\date{7 August 2017}

\begin{document}
\maketitle

\begin{abstract}
Estimation of the direction-of-arrival (DoA) of a speaker in a room is important in many audio signal processing applications. Environments with reverberation that masks the DoA information are particularly challenging. Recently, a DoA estimation method that is robust to reverberation has been developed. This method identifies time-frequency bins dominated by the contribution from the direct path, which carries the correct DoA information. However, its implementation is computationally demanding as it requires frequency smoothing to overcome the effect of coherent early reflections and matrix decomposition to apply the direct-path dominance (DPD) test. In this work, a novel computationally-efficient alternative to the DPD test is proposed, based on the directivity measure for sensor arrays, which requires neither frequency smoothing nor matrix decomposition, and which has been reformulated for sound field directivity with spherical microphone arrays. The paper presents the proposed method and a comparison to previous methods under a range of reverberation and noise conditions. Result demonstrate that the proposed method shows comparable performance to the original method in terms of robustness to reverberation and noise, and is about four times more computationally efficient for the given experiment.

\end{abstract}


\section{Introduction}

The estimation of the direction-of-arrival (DoA) of speakers in reverberant environments using microphone arrays is important in a wide range of applications, including speech enhancement, source separation, robot audition and video conferencing. Methods for DoA estimation have been previously studied extensively. These include beamforming \cite{vanveen1988}, subspace methods such as multiple signal classification (MUSIC) \cite{schmidt1986}, and time-delay of arrival estimation methods \cite{benesty2008}. DoA estimation methods specifically developed for speech signals exploit the non-stationarity and sparsity of speech in the short-time Fourier transform (STFT) domain and enable DoA estimation even for under-determined systems with more sources than microphones \cite{rickard2000,mohan2008,torres2012}. However, in reverberant environments, room reflections mask the direct sound that carries DoA information, thus degrading the DoA estimation performance. 

Recently, a method for DoA estimation of multiple speakers that is robust to reverberation has been developed \cite{nadiri2014}. This method processes the signals in the time-frequency domain, and employs the direct-path dominance (DPD) test to identify time-frequency bins dominated by the direct path. Unlike the earlier coherence test \cite{mohan2008}, the DPD test applies frequency smoothing to significantly reduce the effect of coherent early room reflections. Furthermore, the method is designed for spherical microphone arrays \cite{rafaely2015}, such that DoAs of sources in all directions can be estimated, and frequency smoothing is applied without focusing \cite{wang1985,nadiri2014}. Experimental investigations  validated the robustness of the method to reverberation and its advantages over previous methods \cite{nadiri2014}. The method has been further developed recently, with Gaussian mixture modeling (GMM) applied to the DoA samples to reduce the estimation bias due to the non-normal distribution of the data \cite{rafaely2017icassp}. Although the proposed reverberation-robust method is useful, it is computationally costly, as it requires frequency-smoothing to compute the spatial spectrum matrix for each time-frequency bin, eigen-value decomposition of each of these matrices, and, finally, GMM for data clustering. However, some applications of audio signal processing that employ DoA estimation may be computationally restricted, e.g. robot audition, due to the limited computational resources and the strict real-time requirements. 

This paper proposes a novel alternative to the DPD test which is more computationally efficient than current methods. The new test is based on the directivity measure for sensor arrays \cite{VanTrees}, reformulated for the sound field directivity as measured by the array. This measure has a maximum value for a single-source sound field, and is computed independently for each time-frequency bin. It is shown that the proposed method offers DoA estimation performance that is comparable to that of the recent version of the algorithm \cite{rafaely2017icassp}, while considerably reducing the computation requirements. The paper is structured as follows. An overview of previous methods is outlined in Sec. \ref{sec:DoA_DPD_GMM}, after which the development of the new DPD test is presented in Secs. \ref{sec:directivity_factor} and \ref{sec:DPD_DIR}. A simulation study is detailed in Sec. \ref{sec:simulation_study}, followed by the conclusions in Sec. \ref{sec:conclusions}.

\section{DoA estimation based on DPD test and GMM clustering}
\label{sec:DoA_DPD_GMM}

The current method for speaker localization, recently presented in \cite{nadiri2014} and \cite{rafaely2017icassp} and developed for spherical microphone arrays \cite{rafaely2015}, is outlined in this section. The method is based on the DPD test to identify time-frequency bins that are dominated by the direct sound \cite{nadiri2014}, and GMM for data clustering to reduce estimation bias \cite{rafaely2017icassp}. Consider a spherical microphone array with $Q$ microphones arranged on the surface of a sphere of radius $r$. The sound pressure at the microphones is denoted by $p(k,r,\theta_q,\phi_q)$, with $k$ denoting the wave number and $(\theta_q,\phi_q)$ the spherical coordinates of the angular position of microphone $q$. $\Omega_q\equiv(\theta_q,\phi_q)$ is employed to simplify notation. The sound pressure at the microphones due to $L$ sound sources that surround the array is written in a matrix from as \cite{VanTrees}
\begin{equation}\label{eq:ArrayEqSpace}
\mathbf{p}(k)=\mathbf{V}(k)\mathbf{s}(k)+\mathbf{n}(k),
\end{equation}
with the $Q\times L$ steering matrix $\mathbf{V}(k)$ representing the frequency response from each source to each microphone, the $Q\times 1$ vector $\mathbf{p}(k)$ is given by $\mathbf{p}(k)=[p(k,r,\Omega_1),...,p(k,r,\Omega_Q)]^T$, and the $Q\times1$ vector $\mathbf{n}(k)=[n_1(k),...,n_Q(k)]^T$, represents sensor noise at the microphones. The $L\times1$ vector $\mathbf{s}(k)$ is given by $\mathbf{s}(k)=[s_1(k),...,s_L(k)]^T$, and represents the signal at the $L$ sources. 

For spherical arrays, the steering matrix can be decomposed into frequency-dependent and direction-dependent components, and so Eq. (\ref{eq:ArrayEqSpace}) can be written as \cite{nadiri2014}
\begin{equation}\label{eq:ArrayEqSH}
\mathbf{p}(k)=\mathbf{Y(\Omega)}\mathbf{B}(k)\mathbf{Y}^H(\mathbf{\Psi})\mathbf{s}(k)+\mathbf{n}(k).
\end{equation}
The $Q\times(N+1)^2$ matrix $\mathbf{Y(\Omega)}$ holds the spherical harmonics functions $Y_n^m(\Omega_q)$ of order $n$ and degree $m$, for all $0\le n\le N$ and $-n\le m \le n$. Spherical array design considerations typically lead to an array order $N$ satisfying $(N+1)^2\le Q$ and a frequency range of operation satisfying $kr<N$ (see \cite{rafaely2015} for further details). All orders and degrees in $\mathbf{Y(\Omega)}$ are arranged in a single dimension, leading to a running column index of $n^2+n+m+1$. The $(N+1)^2\times(N+1)^2$ diagonal matrix $\mathbf{B}(k)$ holds radial functions that represent the scattering of a plane wave from a rigid sphere, in the case of a rigid-sphere array, or the elements of the phase response in the case of an open-sphere array \cite{rafaely2015}, and $\mathbf{Y(\Psi})$ is similar to $\mathbf{Y(\Omega)}$, but of size $L\times(N+1)^2$, with $\mathbf{\Psi}$ representing source arrival directions, and $(\cdot)^H$ representing the Hermitian transpose.

Multiplying Eq. (\ref{eq:ArrayEqSH}) from the left by the pseudo-inverse $[\mathbf{Y(\Omega)}]^\dagger$, and the inverse $\mathbf{B}^{-1}(k)$ (regularization may be required at low frequencies or near the nulls of the radial functions \cite{rafaely2015}), leads to plane wave decomposition of the sound field \cite{rafaely2015}, 
\begin{equation}\label{eq:ArrayEqPWD}
\mathbf{a}(k)=\mathbf{Y}^H(\mathbf{\Psi})\mathbf{s}(k)+\mathbf{\tilde{n}}(k),
\end{equation}
where vector $\mathbf{a}(k)$ of size $(N+1)^2\times1$ is given by $\mathbf{a}(k)=[a_{00}(k),...,a_{NN}(k)]^T$. Terms $a_{nm}(k)$ represent the frequency-dependent spherical harmonics coefficients of the plane wave density function measured by the array. Vector $\mathbf{\tilde{n}}(k)$ is the modified sensor noise, given by $\mathbf{\tilde{n}}(k)=\mathbf{B}^{-1}(k)[\mathbf{Y(\Omega)}]^\dagger\mathbf{n}(k)$. 

In the next step, STFT is applied to the plane-wave decomposition signal, leading to

\begin{equation}\label{eq:ArrayEqSTFT}
\mathbf{a}(\tau,\nu)=\mathbf{Y}^H(\mathbf{\Psi})\mathbf{s}(\tau,\nu)+\mathbf{\tilde{n}}(\tau,\nu),
\end{equation}
with $\tau$ denoting the time index and $\nu$ denoting the frequency index. Now, at each time-frequency bin a spatial spectrum matrix is computed, denoted by $\mathbf{R}(\tau,\nu)$, which averages the spatial information over $T$ time bins and $F$ frequency bins in the neighborhood of the selected bin,
\begin{equation}\label{eq:smoothing}
\mathbf{R}(\tau,\nu)=\frac{1}{TF}\sum_{\tau'=\tau}^{\tau+T-1} \sum_{\nu'=\nu}^{\nu+F-1} \mathbf{a}(\tau',\nu')\mathbf{a}^H(\tau',\nu').
\end{equation}
Time averaging is employed to approximate the expectation operation in the computation of the spatial spectrum matrix \cite{VanTrees}. Frequency averaging, also referred to as frequency smoothing, is applied so that the direct sound can be distinguished from coherent reflections. This is important because without frequency smoothing early reflections may bias the DoA estimation, and performance may be significantly degraded under reverberation \cite{nadiri2014}.

The eigen-value decomposition of $\mathbf{R}$ is computed next. If $\mathbf{R}$ is dominated by a single source it is of unit rank, and so the ratio between the first and second singular values is infinite. This motivated the introduction of the DPD test to identify time-frequency bins dominated by a single source, i.e. the direct sound from a speaker \cite{nadiri2014},
\begin{equation}\label{eq:sigma_threshold}
\mathcal{D}=\left\{(\tau,\nu): \frac{\sigma_1(\mathbf{R}(\tau,\nu))}{\sigma_2(\mathbf{R}(\tau,\nu))}\geq \mathcal{TH} \right\},
\end{equation}
where $\sigma_1$, $\sigma_2$ denote the largest and second largest singular values, respectively. $\mathcal{TH}$ is a threshold value, chosen to be sufficiently larger than one to guarantee that $\mathbf{R}$ is dominated by a single singular vector. 

The multiple signal classification (MUSIC) spectrum \cite{schmidt1986}, $P(\Theta,\tau,\nu)$, is computed next for bins that pass the DPD test, i.e. for all $(\tau,\nu)\in\mathcal{D}$, to identify the DoA associated with these bins, 
\begin{equation}\label{eq:MUSIC}
P(\Theta,\tau,\nu)=\frac{1}{\|\mathbf{U_n}^H(\tau,\nu) \mathbf{y^*}(\Theta)
\|^2},\, (\tau,\nu)\in\mathcal{D}
\end{equation}
where $\Theta=(\theta,\phi)$ represents a direction under analysis, and $\mathbf{y}$, of size $(N+1)^2\times 1$, holds the spherical harmonics functions $Y_n^m(\Theta)$ at element number $n^2+n+m+1$. 
With $\mathbf{R}=\mathbf{U\Sigma U}^H$ representing eigen-value decomposition, matrix $\mathbf{U_n}$ is composed of all columns of $\mathbf{U}$ except the first column, which is related to the largest singular value. It therefore represents the noise subspace of $\mathbf{R}$, assuming a single source, and is of dimensions $(N+1)^2 \times [(N+1)^2-1]$. DoA estimation for all bins that passed the DPD test is then computed by
\begin{equation}
\Theta_\mathcal{D}=\left\{\Theta:\arg \max_\Theta P(\Theta,\tau,\nu),\,\forall\, (\tau,\nu)\in\mathcal{D}\right\}.
\end{equation}
Finally, the desired DoA can be computed directly as the mean of the set $\Theta_\mathcal{D}$, although for a more accurate DoA estimation, GMM clustering is applied to $\Theta_\mathcal{D}$ and the mean of the dominant cluster is computed as the desired DoA. This was shown to reduce estimation bias  \cite{rafaely2017icassp}.

It should be noted that the method presented here has been developed for spherical arrays, but can also be applied to any volumetric array that can extract the spherical harmonics coefficients of the sound field \cite{rafaely2015}. The application of the method to other standard arrays, such as uniform linear arrays, may require further study. First, to formulate the DPD test, including frequency-smoothing, for these array configurations, and second, to account for the ambiguities in DoA estimation when linear or planar arrays are positioned in a 3D sound field.

\section{Sound field directivity}
\label{sec:directivity_factor}
In this section a measure for the directivity of a sound field measured by a spherical array is formulated. The new DPD test developed in the following section is based on this measure. The directivity is widely used to characterize the spatial selectivity of array beam patterns \cite{VanTrees,rafaely2015}. In this work it is reformulated to represent the directional property of a sound field surrounding a spherical array. A similar approach was previously presented in \cite{GRS04} to characterize sound fields in rooms. The standard formulation of array directivity is presented first. Given an array with a beam pattern denoted by $\mathcal{B}(\Theta)$, the directivity of the array, denoted by $\mathcal{DIR}$, is given by \cite{VanTrees} 
\begin{equation}\label{eq:directivityArray}
\mathcal{DIR} = \frac{\max_\Theta |\mathcal{B}(\Theta)|^2}{\frac{1}{4\pi}\int_{\mathcal{S}^2} |\mathcal{B}(\Theta)|^2 d\Theta},
\end{equation}
where $\int_{\mathcal{S}^2}d\Theta$ represents an integral over the entire unit sphere surface. To compute the directivity of a sound field, the plane wave density function, $a(\Theta)$, at direction $\Theta$, is first computed from the plane wave coefficients by applying the inverse spherical Fourier transform \cite{rafaely2015},
\begin{equation}\label{eq:pwd}
a(\Theta) = \mathbf{y}^T(\Theta) \mathbf{a}.
\end{equation}
Now, replacing $\mathcal{B}(\Theta)$ with $a(\Theta)$ in Eq. (\ref{eq:directivityArray}), and substituting Eq. (\ref{eq:pwd}), the sound field directivity is formulated as
\begin{equation}\label{eq:directivity}
\mathcal{DIR} = \frac{\max_\Theta |a(\Theta)|^2}{\frac{1}{4\pi}\int_{\mathcal{S}^2} |a(\Theta)|^2 d\Theta} = 
\frac{\max_\Theta |\mathbf{y}^T(\Theta) \mathbf{a}|^2}{\mathbf{a}^H \mathbf{a}}.
\end{equation}
Parseval's theorem for the spherical Fourier transform has been applied to derive the right-hand-side denominator of Eq. (\ref{eq:directivity}). Note that the directivity defined for sound fields in Eq. (\ref{eq:directivity}) and the directivity defined for array beam patterns in Eq. (\ref{eq:directivityArray}), are mathematically equivalent. This is because they both represent the ratio between the maximum and the average magnitude of a function on the sphere. They therefore share the same properties, such as the maximum and minimum values of the directivity, which for spherical arrays are given by \cite{rafaely2015}
\begin{eqnarray}\label{eq:directivity_max_min}
1 \le \mathcal{DIR} \le (N+1)^2.
\end{eqnarray}
The maximum is achieved when vector $\mathbf{a}$ represents a single plane wave, i.e. $a_{nm}=a(k)[Y_n^m(\Psi)]^*$, with $\Psi$ being the wave arrival direction and $a(k)$ being the plane wave amplitude. The minimum is achieved for a constant directivity, e.g. in an ideal diffuse sound field \cite{rafaely2015}.

\section{DPD test based on sound field directivity}
\label{sec:DPD_DIR}

A new alternative to the DPD test is proposed in this section, based on the sound field directivity defined in Eq. (\ref{eq:directivity}). The directivity is computed for each time-frequency bin from the plane-wave decomposition vector $\mathbf{a}$ using Eq. (\ref{eq:ArrayEqSTFT}),
\begin{equation}\label{eq:dir_samp}
\mathcal{DIR}(\tau,\nu) = \frac{\max_\Theta |\mathbf{y}^T(\Theta) \mathbf{a}(\tau,\nu)|^2}{\mathbf{a}^H(\tau,\nu) \mathbf{a}(\tau,\nu)}.
\end{equation}
Recall that the maximum directivity, $(N+1)^2$, is achieved for a sound field composed of a single plane wave, representing a single source. Now, because the DPD test aims to identify time-frequency bins dominated by a single source, the direcitivty can be used as an alternative measure for the DPD test,
\begin{equation}\label{eq:dir_threshold}
\mathcal{D}=\left\{(\tau,\nu): \mathcal{DIR}(\tau,\nu) \geq \alpha (N+1)^2 \right\},
\end{equation}
where $\frac{1}{(N+1)^2}\le \alpha\le 1$ is used as a threshold parameter. The choice of $\alpha$ close to 1 leads to the selection of bins with the highest directivity, i.e. representing a single plane-wave sound field. Note that any sound field that is not composed of a single plane wave will have a directivity that is lower than the maximum, whether the sources that compose this sound field are coherent or incoherent. This is the result of the orthogonality property of the spherical harmonics \cite{rafaely2015}. The inner product between $\mathbf{y}(\Theta)$ for a given $\Theta$ and $\mathbf{a}(\tau,\nu)$ in Eq. (\ref{eq:dir_samp}) will be maximal when they are equal (but complex conjugate), i.e. when vector $\mathbf{a}$ represents a single plane wave arriving from $\Theta$. Therefore, by using the direcitivity as a measure for identifying time-frequency bins dominated by a single source, neither frequency smoothing nor matrix decomposition are required (because there is no need to construct the correlation matrix) - only a single computation of the directivity per bin. Denoting $l=(N+1)^2$, the length of vector $\mathbf{a}$, the computational complexity of the DPD test with matrix decomposition is of order $\mathcal{O}(l^3)$ \cite{nadiri2014,demmel1997}, while for the proposed DPD test the computational complexity is of order $\mathcal{O}(l)$, due to the vector-based computations. 

The estimated DoA for each bin that passed the DPD test is simply the angle at which $|a(\Theta)|^2$ is maximum,
\begin{equation}\label{eq:theta_dir}
\Theta_\mathcal{D}=\left\{\Theta:\arg\max_\Theta |\mathbf{y}^T(\Theta)\mathbf{a}(\tau,\nu)|^2,\,\forall\, (\tau,\nu)\in\mathcal{D}\right\}, 
\end{equation}
which was already computed in Eq. (\ref{eq:dir_samp}). Finally, similar to the method in Sec. \ref{sec:DoA_DPD_GMM}, the desired DoA can be computed as the mean of the set $\Theta_\mathcal{D}$, or as the mean of the dominant cluster after GMM clustering is applied to $\Theta_\mathcal{D}$.
Note that although a single speaker formulation is developed here for simplicity, it can be extended to several speakers by clustering DoAs of individual speakers, in a way similar to that described in \cite{nadiri2014}.

With the aim of highlighting the differences in computation complexity between the previously proposed method, denoted \textsf{THR-GMM} \cite{nadiri2014,rafaely2017icassp} and the newly proposed method, denoted \textsf{DIR-GMM}, these differences are outlined in Table \ref{tab:computations}. The table details the various stages, i.e. time-frequency averaging, eigen-value decomposition, grid search, and GMM, showing whether the methods require each stage and the computations required, with a reference to the corresponding equation. The table clearly shows the advantages of the proposed method - neither eigen-value decomposition nor time-frequency averaging are required, and the direction-finding grid search is simpler.
\begin{table}[ht]
\centering
\begin{tabular}{ l|c c c c}
  \textbf{Method} & \textbf{TF averaging} & \textbf{EVD} & \textbf{Grid search} & \textbf{GMM} \\
 \hline
  \textsf{THR-GMM} & $T\times F$ (\ref{eq:smoothing}) & $\mathcal{O}\left((N+1)^6\right)$ & $(N+1)^2 \times [(N+1)^2-1]$ (\ref{eq:MUSIC}) & $\checkmark$ \\
  \hline
  \textsf{DIR-GMM} & \ding{55} & \ding{55} & $(N+1)^2$ (\ref{eq:theta_dir}) & $\checkmark$\\
  \hline
\end{tabular}
\caption{Comparison of computation complexity of the two methods: \textbf{TF averaging} is time-frequency averaging (summations) per time-frequency bin; \textbf{EVD} is eigen-value decomposition per bin; \textbf{Grid search} is employed to estimate the DoA for each bin, denoting the number of multiplications per grid angle per bin; and $\checkmark$ denotes that a similar \textbf{GMM} is required in both methods.\label{tab:computations}}
\end{table}

\section{Simulation study}
\label{sec:simulation_study}

The previously proposed methods and the newly proposed method for the DPD test are studied in this section using a simulation example, and compared to the standard generalized cross-correlation, phase transform (GCC-PHAT) method \cite{benesty2008}. The aim of the study is to provide insight into the behavior of the various tests, and to evaluate their performance when integrated into a DoA estimation process. In the simulation, a single speaker is positioned in a room of dimensions $8\times5\times3\,$m, under various reverberation conditions, ranging from anechoic to a reverberation time of $1\,$s, and a distance from the speaker to the microphone array which is up to $3$ times the critical distance. The impulse responses from the speaker, modeled as a point source, to an open-sphere spherical microphone array \cite{rafaely2015} with 32 microphones of order $N=3$, positioned $1.8\,$m away from the source, are simulated using the image method \cite{allen1979}. Microphone signals are produced by convolving the room impulse responses with speech from the TIMIT database \cite{garofolo1993}, sampled at  $16\,$kHz. An FFT size of 512 samples was used in the STFT analysis, employing a Hann window with $50\%$ overlap. Sensor noise is introduced at a signal-to-noise ratio (SNR) of $40\,$dB to simulate the effect of microphone amplifier noise.   

\begin{figure}[htb]
\centering
\includegraphics[width=0.8\columnwidth]{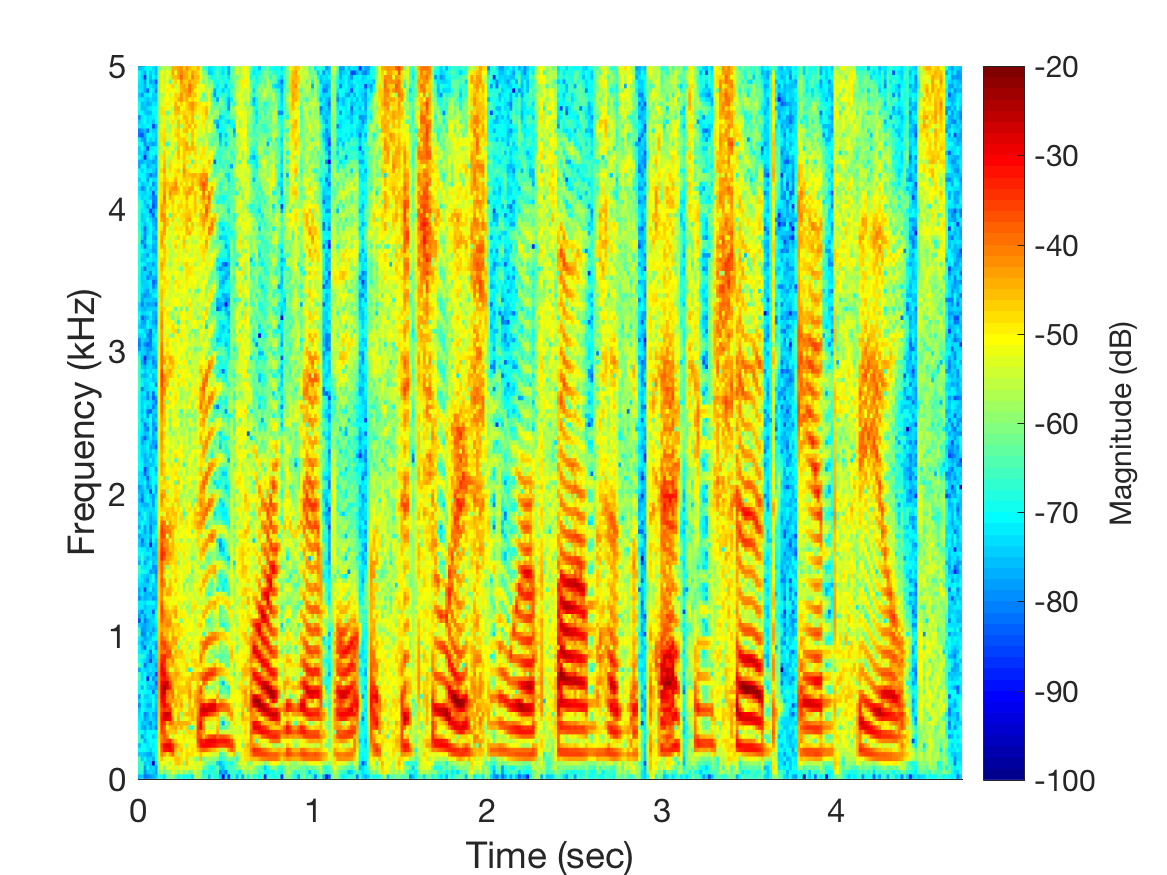}
\includegraphics[width=0.8\columnwidth]{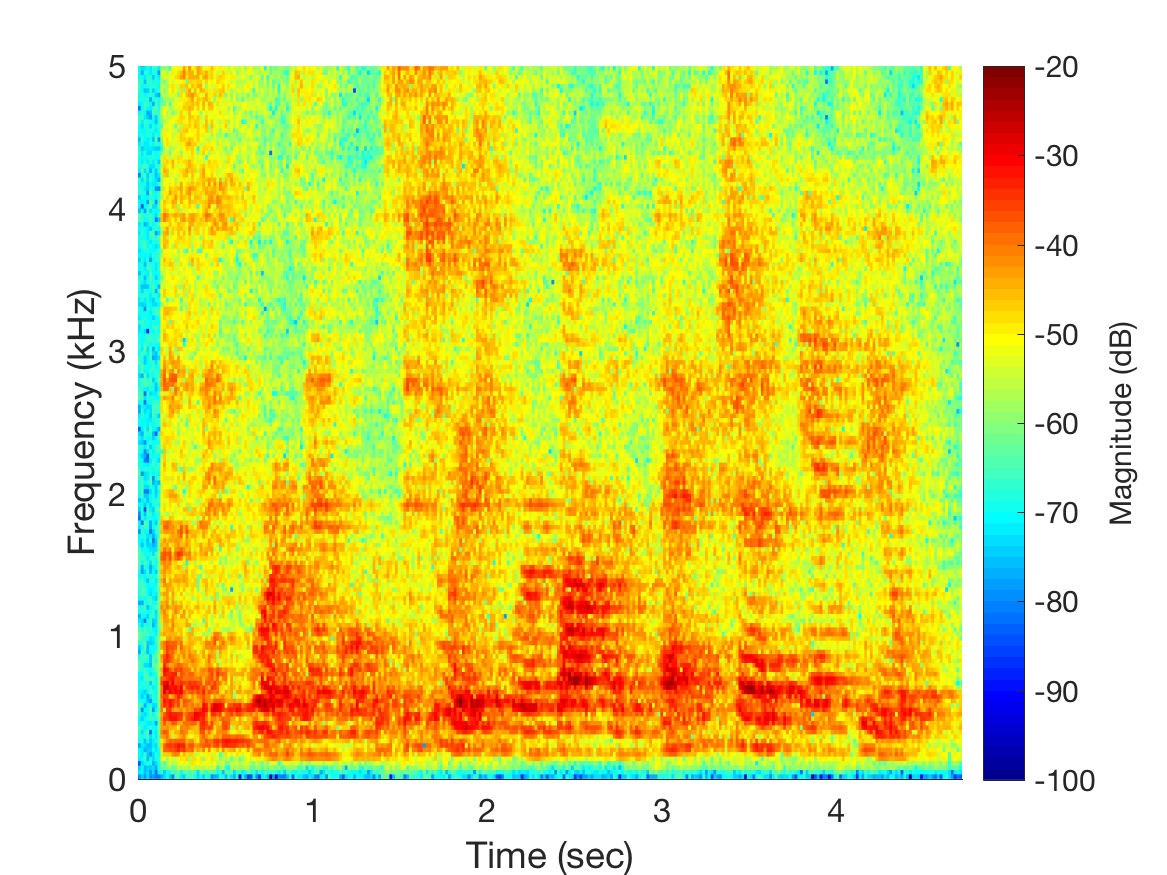}
\caption{Spectrogram of clean speech (upper) and reverberant speech ($T_{60}=1\,$s) as measured by one of the array microphones, including sensor noise (lower)}.\label{fig:speech}
\end{figure}

Figure \ref{fig:speech} presents spectrograms of the clean and reverberant speech signals, as measured by one of the array microphones. 
The figure clearly shows the time-smearing effect of reverberation on the speech signal. Next, the DPD test is applied using two different methods, as follows.

\begin{itemize}
\item[(i)] \textsf{THR-GMM} (the method of \cite{rafaely2017icassp}): $\mathcal{D}$ is computed using Eq. (\ref{eq:sigma_threshold}) with $\mathcal{TH}=2$; matrix $\mathbf{R}$ is computed using Eq. (\ref{eq:smoothing}) with $T=2$ and $F=15$; $\Theta_\mathcal{D}$ is computed using MUSIC, as in Eq. (\ref{eq:MUSIC}); GMM with 3 Gaussians and a diagonal covariance matrix is applied to select the subset in $\Theta_\mathcal{D}$ belonging to the dominant Gaussian (the one with the largest mixing weight); and the desired DoA is computed as the mean of this subset of $\Theta_\mathcal{D}$.
\item[(ii)] \textsf{DIR-GMM} (the new method): $\mathcal{D}$ is computed using Eq. (\ref{eq:dir_threshold}) with $\alpha=0.4$; $\Theta_\mathcal{D}$ is computed using Eq. (\ref{eq:theta_dir}); GMM with 3 Gaussians and a diagonal covariance matrix is applied to select the subset in $\Theta_\mathcal{D}$ belonging to the dominant Gaussian; and the desired DoA is computed as the mean of this subset of $\Theta_\mathcal{D}$.
\end{itemize}

In addition, the methods are compared to the well established method for DoA estimation, GCC-PHAT  \cite{benesty2008}, realized in the commercially available Phased Array System Toolbox in MATLAB version R2016B.

\subsection{Comparative analysis of the DPD tests}

\begin{figure}
\centering
\includegraphics[width=0.78\columnwidth]{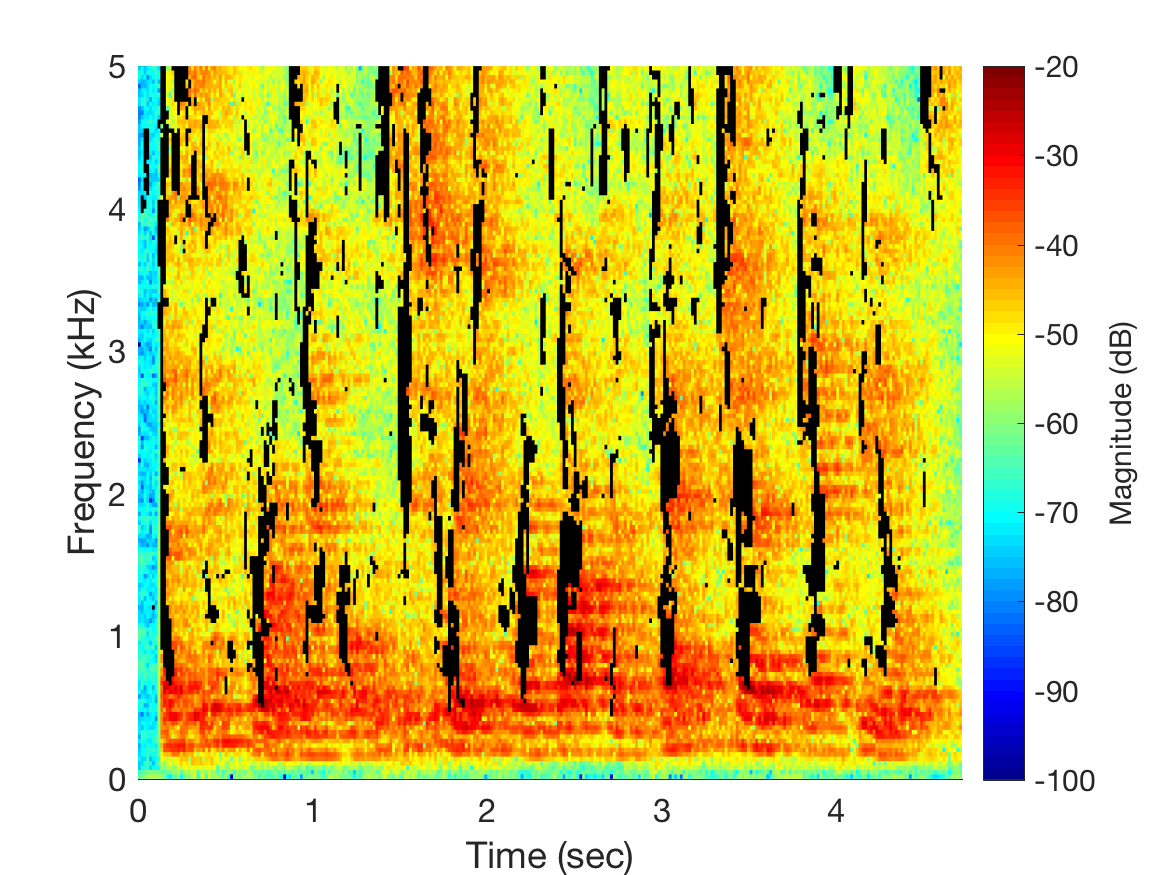}
\includegraphics[width=0.78\columnwidth]{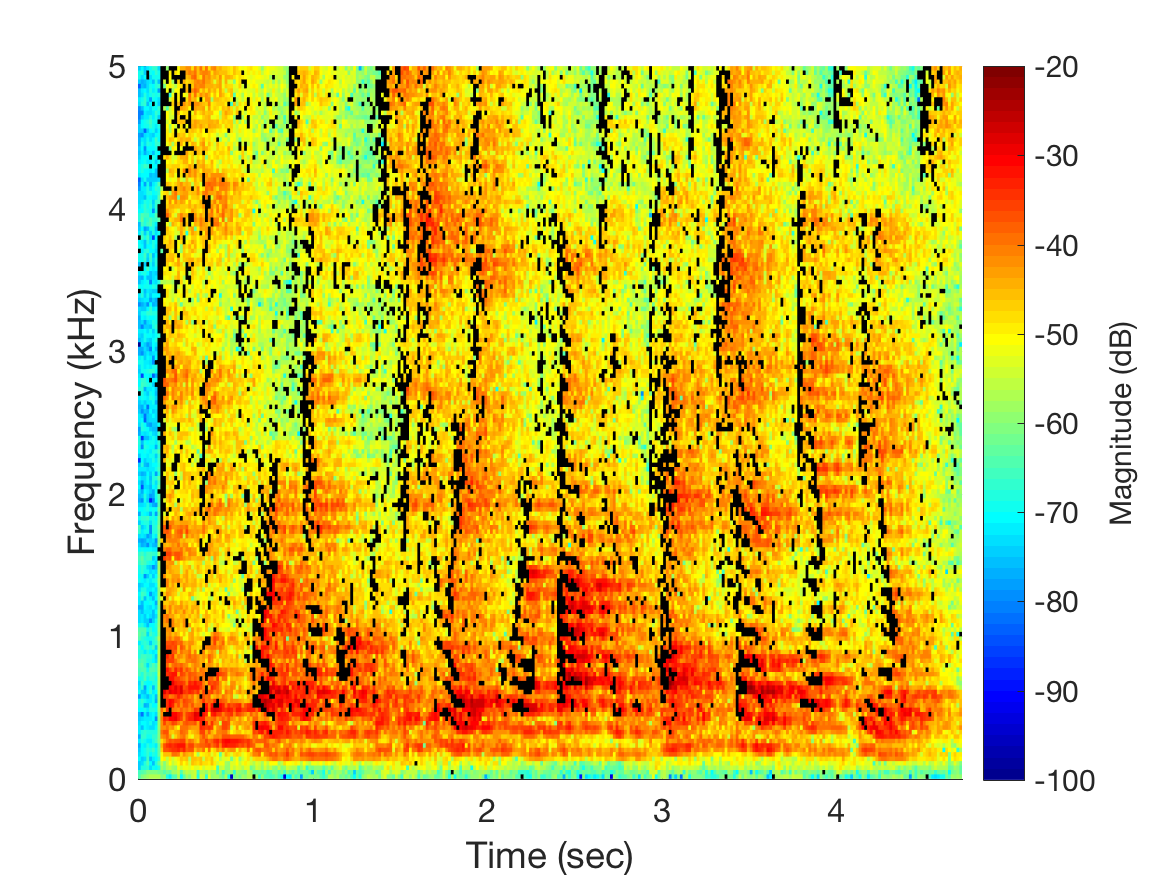}
\caption{Time-frequency maps for bins that passed the DPD test using the \textsf{THR-GMM} method (upper), with $\mathcal{TH}=2$, and the \textsf{DIR-GMM} method, with $\alpha=0.4$ (lower), presented over the spectrograms of the reverberant speech.}\label{fig:DPDmaps}
\end{figure}

In this section an analysis of the bins that passed the DPD tests is presented. Figure \ref{fig:DPDmaps} present time-frequency maps of these bins, i.e. $(\tau,\nu)\in\mathcal{D}$, for the two methods. These bins are colored in black, while the reverberant speech spectrogram is presented as before. The conditions of this simulation included a reverberation time ($T_{60}$) of $1\,$s, and sensor noise with SNR of $40\,$dB.  The figures illustrate that time-frequency bins passing the DPD tests are predominantly at the onsets of the speech signal. This is expected, as discussed in \cite{nadiri2014} - speech sounds that arrive after a short pause in speech are dominated by the first-arriving direct sound. In contrast, bins that follow are contaminated by early room reflections and reverberation, and are no longer identified as direct sound. Figure \ref{fig:DPDmaps} shows that both the \textsf{THR-GMM} and \textsf{DIR-GMM} methods seem to identify bin in similar regions in the spectrogram, although the \textsf{THR-GMM} seems to produce bins in clusters, while \textsf{DIR-GMM} produces more isolated or sparse bins. This can be attributed to the averaging process in the \textsf{THR-GMM} methods, which has a smoothing effect on the selected bins. In contrast, \textsf{DIR-GMM} operates on individual bins, therefore leading to the more sparse bin selection.
Although this analysis provided some insight into the behavior of the two DPD tests, also clearly illustrating the distributions of the selected bins on speech spectrograms, it only inspected the methods under a single set of conditions, and only presented spectrogram maps and not DoA estimation performance. The following study presents DoA estimation performance under various acoustic conditions of reverberation and noise.

\subsection{DoA estimation study}

\begin{figure}
\centering
\includegraphics[width=0.8\columnwidth]{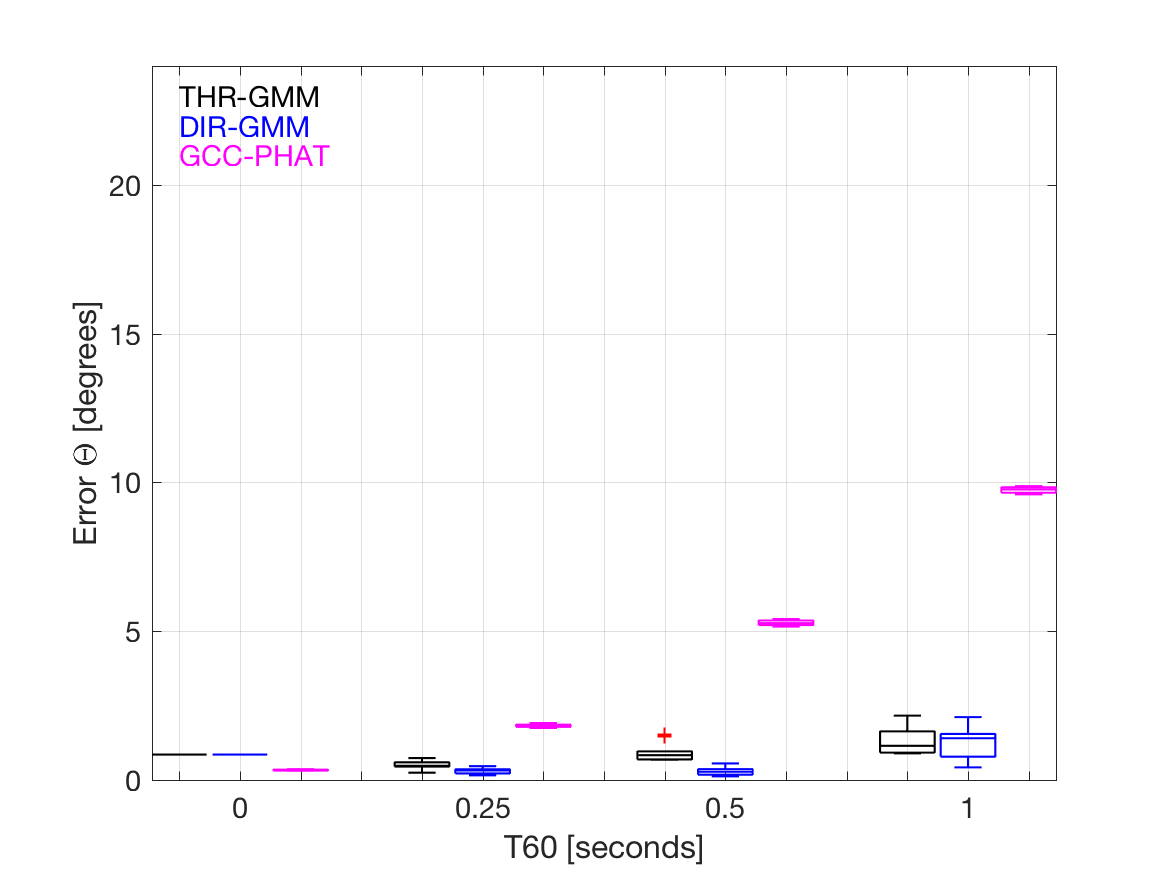}
\caption{\textcolor{blue}{DoA estimation error with the three methods (\textsf{THR-GMM}, \textsf{DIR-GMM} and GCC-PHAT), as a function of reverberation time. Sensor noise and diffuse noise are fixed at $40\,$dB SNR. The figure shows box plots, with median values, first and third quartiles, minimum and maximum values, and any samples identified as outliers (\textcolor{red}{+}). \label{fig:T60}}}
\end{figure}

In this subsection the performance of the two methods is evaluated in the context of DoA estimation. After identifying the set of bins that passed the two DPD tests, the desired DoA is estimated as detailed above. However, to ensure that results do not reflect only a very specific condition, DoA estimation was performed under a wider set of conditions: 

\begin{enumerate}
\item[(i)] By changing the reflection coefficients of the room walls, leading to four room reverberation conditions, i.e. anechoic (denoted as $T_{60}=0\,$s), and $T_{60}=0.25\,$s, $0.5\,$s and $1\,$s,
\item[(ii)] By adding diffuse noise, with SNR ranging from $40\,$dB to $0\,$dB. The diffuse noise was added directly as an uncorrelated noise to the spherical harmonics coefficients of the measured sound field, vector $\mathbf{a}$ in Eq. (\ref{eq:ArrayEqPWD}) \cite{rafaely2015},
\item[(iii)] By including 5 different speakers.
\end{enumerate}
{Figure \ref{fig:T60} presents the first set of results of DoA estimation errors. For each reverberation time condition, the sensor noise and diffuse noise were fixed at an SNR of $40\,$dB. \textcolor{blue}{Box plots are presented for each condition, representing the statistics} for 10 realizations of the sensor noise and diffuse noise, and a selection of one speaker from the speakers set. The figures show that with no reverberation, i.e. $T60=0\,$s, all methods perform well, with errors below $1^\circ$. In fact, the GCC-PHAT method outperform the DPD test based methods under anechoic conditions. 
As reverberation time increases, the \textsf{THR-GMM} and \textsf{DIR-GMM} methods maintain good performance, which slightly degrades at $T_{60}=1\,$s. These results reinforce previous results \cite{nadiri2014,rafaely2017icassp} and demonstrate that the DPD test based methods work well under reverberation. In contrast, GCC-PHAT, which is not designed specifically to be robust to reverberation, shows increasing error levels with reverberation, reaching $10^\circ$ at $T_{60}=1\,$s.}

\begin{figure}
\centering
\includegraphics[width=0.8\columnwidth]{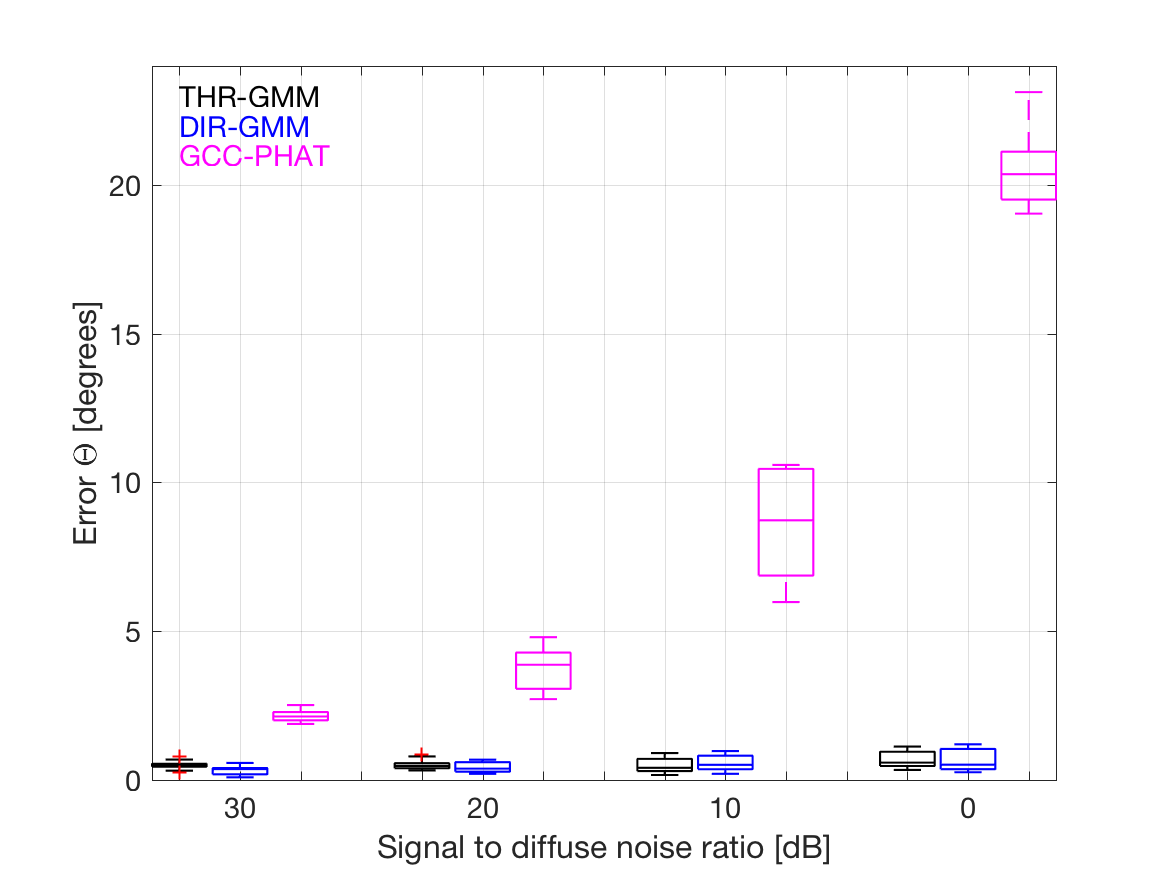}
\caption{\textcolor{blue}{DoA estimation error with the three methods (\textsf{THR-GMM}, \textsf{DIR-GMM} and GCC-PHAT), as a function of SNR with respect to a diffuse noise field. Sensor noise is fixed at $40\,$dB SNR. The figure shows box plots, with median values, first and third quartiles, minimum and maximum values, and any samples identified as outliers (\textcolor{red}{+}).\label{fig:Diffuse}}}
\end{figure}

Figure \ref{fig:Diffuse} presents the second set of results. Here, DoA estimation errors are investigated as a function of diffuse noise SNR. A room with a reverberation time of $T_{60}=0.25\,$s was employed in this case. The \textsf{THR-GMM} and the \textsf{DIR-GMM} methods exhibit similarly good performance, with an error that is slightly increasing with decreasing SNR, but which \textcolor{blue}{is mostly below} $1^\circ$. On the other hand, the performance of the GCC-PHAT method significantly degrades with noise level. This shows another advantage of the proposed method - robustness to noise, which has not been previously investigated in any significant manner. This behavior is not surprising. Although the method was primarily designed to be robust to reverberation, its inherent nature makes it robust to noise as well. Time-frequency bins that are contaminated by noise, will score low on both DPD tests, and will therefore not be selected for DoA estimation.

This simulation study demonstrated the robustness of both the \textsf{THR-GMM} and \textsf{DIR-GMM} methods to reverberation and noise. In particular, the proposed \textsf{DIR-GMM} method demonstrated comparable performance to the \textsf{THR-GMM} method over a wide range on conditions. As discussed in Section \ref{sec:DPD_DIR}, the proposed \textsf{DIR-GMM} method is more computationally efficient, because neither averaging of time-frequency bins nor eigen-value decomposition are required. To quantify this computational advantage with respect to the simulation study in this section, the running times of the \textsf{THR-GMM} and \textsf{DIR-GMM} methods were compared. Both methods were realized in MATLAB version R2016b, and the average running time on a Macbook Pro laptop with $8\,$GB of RAM and $2.7\,$GHz Intel i5 processor, under all conditions for the entire speech segment (about $5\,$s) were compared. Average running times are presented in Table \ref{tab:times}. It is clear from the table, that on average, the \textsf{DIR-GMM} method is more than $4$ times faster than the \textsf{THR-GMM}. It is also interesting to note that the GCC-PHAT method is faster than the \textsf{THR-GMM} method but slower than the \textsf{DIR-GMM} method.
\begin{table}
\centering
\begin{tabular}[ht]{l|c}
\textbf{Method} & \textbf{Time} \\
\hline
\textsf{THR-GMM} & $62.0\,$s \\
\textsf{DIR-GMM} & $13.5\,$s \\
GCC-PHAT & $26.9\,$s \\
\hline\hline
\end{tabular}
\caption{Running times for the three methods under MATLAB version 2016b and a Macbook Pro with $8\,$GB of RAM, for about $5\,$s of speech, averaged over all conditions.\label{tab:times}}
\end{table}

To summarize, the proposed \textsf{DIR-GMM} method for the DPD test provided DoA estimation accuracy comparable to that of the \textsf{THR-GMM} method under a wide range of conditions, while demonstrating significant improvement in computational efficiency. Although the methods were studied under challenging conditions, an experimental study of the proposed method may be desirable. This, however, is beyond the scope of this paper, and is therefore proposed for a future study.

\section{Conclusions}\label{sec:conclusions}

A new computationally efficient method for the direct-path dominance test has been proposed in this paper. The method is based on the measure of directivity of the sound field, and can facilitate DoA estimation with high accuracy at low computational complexity. The method may therefore be particularly useful in applications that require accurate speaker localization under reverberation and noise, but may have computation restrictions, such as robot audition and mobile communication. 

% BiBTeX files. The IEEEbib.bst bibliography
\bibliographystyle{IEEEbib}
\bibliography{refs,main}

\begin{thebibliography}{10}

\bibitem{vanveen1988}
B.~D.~Van Veen and K.~M. Buckley,
\newblock ``Beamforming: A versatile approach to spatial filtering,''
\newblock {\em IEEE ASSP Mag.}, vol. 5, no. 2, pp. 4--24, 1988.

\bibitem{schmidt1986}
R.~Schmidt,
\newblock ``Multiple emitter location and signal parameter estimation,''
\newblock {\em IEEE Trans. Antennas Propag.}, vol. 34, no. 3, pp. 276--280,
  march 1986.

\bibitem{benesty2008}
J.~Benesty, J.~Chen, and Y.~Huang,
\newblock {\em Microphone Array Signal Processing},
\newblock Springer, 1 edition, 2008.

\bibitem{rickard2000}
S.~Rickard and F.~Dietrich,
\newblock ``{DOA} estimation of many {W}-disjoint orthogonal sources from two
  mixtures using duet,''
\newblock in {\em Statistical Signal and Array Processing, 2000. Proceedings of
  the Tenth IEEE Workshop on}, 2000, pp. 311--314.

\bibitem{mohan2008}
S.~Mohan, M.~E. Lockwood, M.~L. Kramer, and D.~L. Jones,
\newblock ``Localization of multiple acoustic sources with small arrays using a
  coherence test,''
\newblock {\em J. Acoust. Soc. Am.}, vol. 123, no. 4, pp. 2136--2147, 2008.

\bibitem{torres2012}
A.~M. Torres, M.~Cobos, B.~Pueo, and J.~J. Lopez,
\newblock ``Robust acoustic source localization based on modal beamforming and
  time-frequency processing using circular microphone arrays,''
\newblock {\em J. Acoust. Soc. Am.}, vol. 132, no. 3, pp. 1511--1520, 2012.

\bibitem{nadiri2014}
O.~Nadiri and B.~Rafaely,
\newblock ``Localization of multiple speakers under high reverberation using a
  spherical microphone array and the direct-path dominance test,''
\newblock {\em IEEE Trans. Audio, Speech, Language Process.}, vol. 22, no. 10,
  pp. 1494--1505, October 2014.

\bibitem{rafaely2015}
B.~Rafaely,
\newblock {\em Fundamentals of Spherical Array Processing},
\newblock Springer-Verlag, Berlin, first edition, 2015.

\bibitem{wang1985}
H.~Wang and M.~Kaveh,
\newblock ``Coherent signal-subspace processing for the detection and
  estimation of angles of arrival of multiple wideband sources,''
\newblock {\em IEEE Trans. Acoust., Speech, Signal Process.}, vol. 33, no. 4,
  pp. 823--831, August 1985.

\bibitem{rafaely2017icassp}
B.~Rafaely and D.~Kolossa,
\newblock ``Speaker localization in reverberant rooms based on direct path
  dominance test statistics,''
\newblock in {\em IEEE International Conference on Acoustics, Speech, and
  Signal Processing (ICASSP 2017), Accepted for publication}, 2017.

\bibitem{VanTrees}
H.~L. Van-Trees,
\newblock {\em Optimum Array Processing (Detection, Estimation, and Modulation
  Theory, Part IV)},
\newblock Wiley-Interscience, 1 edition, 2002.

\bibitem{GRS04}
B.N. Gover, J.G. Ryan, and M.R. Stinson,
\newblock ``Measurement of directional properties of reverberant sound fields
  in rooms using a spherical microphone array,''
\newblock {\em J. Acoust. Soc. Am.}, vol. 116, pp. 2138--2148, 2004.

\bibitem{demmel1997}
J.~W. Demmel,
\newblock {\em Applied Numerical Linear Algebra},
\newblock Society for Industrial and Applied Mathematics, 1997.

\bibitem{allen1979}
J.~B. Allen and D.~A. Berkley,
\newblock ``Image method for efficiently simulating small-room acoustics,''
\newblock {\em J. Acoust. Soc. Am.}, vol. 65, no. 4, pp. 943--950, 1979.

\bibitem{garofolo1993}
J.~S. Garofolo, L.~F. Lamel, W.~M. Fisher, J.~G. Fiscus, D.~S. Pallet, and
  N.~S. Dahlgren,
\newblock ``{DAPRA TIMIT} acoustic-phonetic continuous speech corpus,''
\newblock CDROM, 1993, US National Institute of Standards and Technology.

\end{thebibliography}
\end{document}